\documentclass[a4paper]{spie}
\pdfoutput=1
\usepackage[utf8]{inputenc} 
\usepackage{url}
\usepackage{graphicx}
\usepackage{algorithmic}
\usepackage{algorithm}
\usepackage{amsmath}
\usepackage{natbib}

\newcommand{\sSeq}[1]{\texttt{#1}}

\title{Assembly of repetitive regions using next-generation sequencing data}

\author{Robert M. Nowak
  \skiplinehalf
  Institute of Electronic Systems, Warsaw University of Technology; Nowowiejska 15/19, 00-665, Warsaw, Poland
}

\authorinfo{Further author information: (Send correspondence to Robert Nowak)\\
  Robert Nowak: Tel: +48 22 2347718; E-mail: r.m.nowak@elka.pw.edu.pl}

\setcitestyle{numbers,square}
\begin{document}
\maketitle

\begin{abstract}
High read depth can be used to assemble short sequence repeats.
The existing genome assemblers fail in repetitive regions of longer than average read.

I propose a~new algorithm for a~DNA assembly which uses the relative frequency of reads to properly reconstruct repetitive sequences.
The mathematical model shows the upper limits of accuracy of the~results as a~function of read coverage.
For high coverage, the estimation error depends linearly on repetitive sequence length and inversely proportional to the sequencing coverage.
The algorithm requires high read depth, provided by the next-generation sequencers and could use the existing data.
The tests on errorless reads, generated in silico from several model genomes, pointed the properly reconstructed repetitive sequences,
where existing assemblers fail.

The C++ sources, the Python scripts and the additional data are available at \url{http://dnaasm.sourceforge.org}.
\end{abstract}


\section{Introduction}

Next-generation sequencing (NGS) dramatically reduced the cost of producing genome sequences \cite{shendure2008next}.
Therefore, we observe exponential increase of sequencing data \cite{pagani2012genomes}.
The whole-genome shotgun method is the most popular sequencing technique,
where the computer programs called genome assemblers reconstruct a DNA sequence up to chromosome length.
The genome assembly is a~challenging task for computer science due to a~huge volume and complexity of input data produced by NGS.
The huge volume of data results from both higher throughput and higher over-sampling.
Computer programs use the de~Bruijn graphs \cite{pevzner2001eulerian,myers2005fragment} as well as greedy extensions of overlap-consensus-layout graphs \cite{miller2010assembly} to process the volume of data.

Currently more than 50 genome assemblers are available \cite{zhang2011practical,earl2011assemblathon,bradnam2013assemblathon,salzberg2012gage},
but the assembly products are incomplete due to the repetitive regions, the uncovered areas and the sequencing errors.
The feasibility of assembly with short reads generated from completely sequenced genomes \cite{kingsford2010assembly}
shows that there is still room for better algorithms.

The short sequence repeats (SSR) are infrequent in sequences coding proteins, therefore transcriptome analysis use genome sequences without properly restored SSR.
However, SSR occur in large quantities in eukaryiotic \cite{cox1997characteristic} and prokaryitic cells \cite{van1998short},
mainly in extragenic and regulatory regions and these regions are used to study genetic variations between individuals.
Older techniques based on micro-array or electrophoresis have been replaced with
the NGS data used to detect such variations \cite{cao2013inferring,xie2009cnv,yoon2009sensitive},
when the reference genome is available.

\begin{figure}[!htb]
 \scalebox{0.9}{\includegraphics{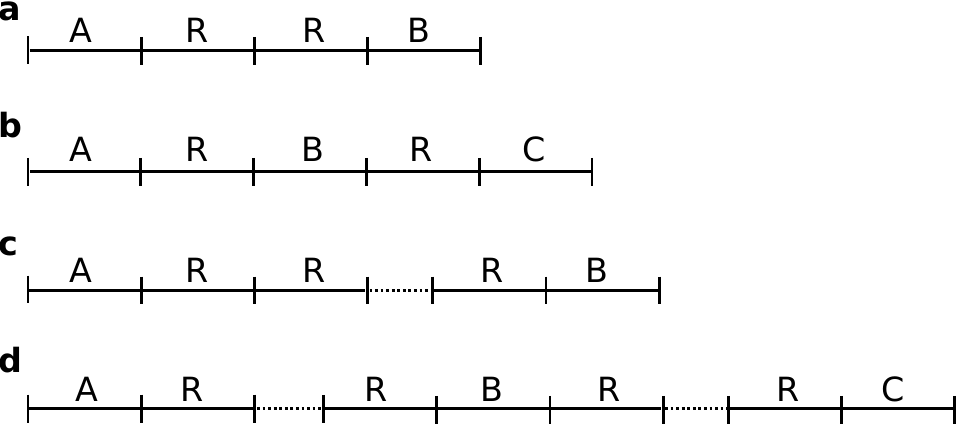}}

  \label{figRepetitiveSolvable}
  \caption{
    SSR, denoted by R, surrounded by unique sequences, denoted by A, B, C, in genome fragment.
    Case \textbf{a} and \textbf{b} are handled by existing sequence assemblers, case \textbf{c} is properly solved by the presented algorithm,
    case \textbf{d} is insolvable. Each A, B, C, R are longer than de~Bruijn graph dimension.
  }
\end{figure}

In the presented approach I propose a~new algorithm to retrieve the length of an~repetitive section using short reads,
designed for \textit{de~novo} assembly of NGS data.
This algorithm estimates SSR length from the coverage statistics
and it is able to properly assemble consecutive repeats, as depicted in Fig.~\ref{figRepetitiveSolvable}.

To my knowledge, only the Euler-SR assembler\cite{chaisson2008short} handles consecutive repeats of longer than average read or de~Bruijn graph dimension.
It constructs the assembly as a~path that traverses the repeat twice, therefore underestimates the copy number.
The other assemblers skip such SSR.

The paper is organized as follows:
Section~\ref{secApproach} describes the new algorithm and the mathematical model used to calculate the accuracy of SSR length estimation.
Section~\ref{secMethods} shows the numerical experiments on \textit{in silico} generated data.
Finally, Section~\ref{secDiscussion} presents the proposals for extensions, the protocol of processing the existing data and the conclusions.

\section{Approach}
\label{secApproach}

\subsection*{Algorithm}

The algorithm uses a~k-dimensional weighted de~Bruijn multigraph $G(V,E)$, called A-Bruijn graph \cite{pevzner2004novo},
where $V$ is a~set of vertices, $E$ is a~set of edges.
The edge $e(u,v)$ represents the sequence $s_{0}s_{1}...s_{k-1}$,
the vertex $u$, the source of edge $e$, represents the sequence $s_{0}...s_{k-2}$, the vertex $v$, the target of $e$, represents $s_{1}...s_{k-1}$.
The edge weight $w$ may be understood as the~number of the~parallel edges between the source and the target vertices
and it depicts how many times the edge should be used to produce an output path.

The algorithm is built of three steps:
the graph construction from reads, the edge weight normalization and the output generation.

Alg.~\ref{algGraphConstruct} constructs an A-Bruijn graph $G$ from a set of reads $R$.
Every sub-string of length $k$ from $R$ creates an edge in $G$.

\begin{algorithm}
  \caption{A-Bruijn graph construction algorithm}
  \label{algGraphConstruct}
  \begin{algorithmic}
    \REQUIRE $R$ collection of reads
    \STATE $G \gets \emptyset$
    \FORALL{$r \in R : |r| \ge k$}
    \FORALL{$i : 0 \le i \le |r|-k$}
    \STATE $u \gets r_{i}...r_{i+k-2}$, $v \gets r_{i+1}...r_{i+k-1}$
    \STATE $G.add(edge(u,v))$ \COMMENT{increase edge's weight if exists otherwise add new edge to $G$ with $w=1$}
    \ENDFOR
    \ENDFOR
  \end{algorithmic}
\end{algorithm}

An SSR is a sequence $S=m_{0}...m_{d-1}m_{0}...m_{(n-1)\bmod d}$, with of length $n$, $|S| = n$.
$S$ is built of a repeating motif $m=m_{0}m_{1}...m_{d-1}$, $|m|=d$, $d \le \frac{n}{2}$.
Such SSR create whirls \cite{chaisson2008short}, when $n \ge 2(k-1)$.
Some whirls are shown in Tab.~\ref{tabWhirlsBruijn}.

\begin{table}[!htb]
  \centering

  \begin{tabular}{|c|c|c|c|} \hline
    ~ & $d=1$ & $d=2$ & $d=3$  \\ \hline
    $
    \begin{array}{l}
      n -k \equiv 0\\
      \pmod{d}
    \end{array}
    $
    &
    &  \scalebox{0.7}{\includegraphics{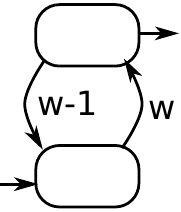}}
    &  \scalebox{0.7}{\includegraphics{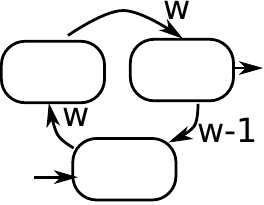}}    \\ \cline{1-1} \cline{3-4}

    $
    \begin{array}{l}
      n -k \equiv 1\\
      \pmod{d}
    \end{array}
    $
    &
    &
    &  \scalebox{0.7}{\includegraphics{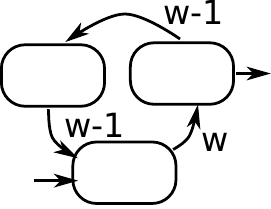}}    \\ \cline{1-1} \cline{4-4}

    $
    \begin{array}{l}
      n -k \equiv d-1\\
      \pmod{d}
    \end{array}
    $
    & \scalebox{0.7}{\includegraphics{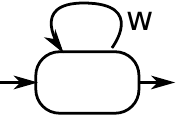}}
    & \scalebox{0.7}{\includegraphics{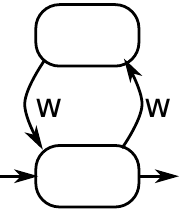}}
    &  \scalebox{0.7}{\includegraphics{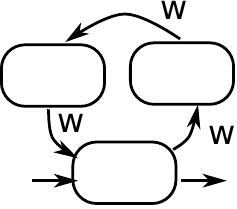}}    \\ \hline

  \end{tabular}

  \caption{Examples of whirls in an~A-Bruijn graph, $w$ is edge weight, $k$ is graph dimension, $d$ is motif length, $n$ is repetitive sequence length,
    $n \ge 2(k - 1)$.}
  \label{tabWhirlsBruijn}
\end{table}

The second step of the algorithm, edge normalization, is a new approach to genome assembly.
Eq.~\ref{eqRounding} converts the edge weight $w$ into $w'$,
where $c$ is sequencing coverage and $L$ is read average length.
The $w'$ may be understood as edge coverage, because the sequence with length $L$ creates $L-k+1$ fragments of length $k$ in Alg.~\ref{algGraphConstruct}.

\begin{equation}
  w'=\lfloor \frac{k}{c(L-k+1)}w + 0.5 \rfloor, \mbox{~where~} L \ge k
  \label{eqRounding}
\end{equation}

The normalization reduces errors, assuming that reads spread uniformly over the sequenced genome.
The fragments that occur less frequently than $\frac{c(L-k+1)}{2k}$ are removed from A-Bruijn graph.
The normalization plays a similar role to rejecting the sequences that occurs less frequently than predetermined threshold,
which is used in other sequence assemblers.
Moreover, it provides the relative frequency of edges, used for a proper SSR assembly.

The final step of the algorithm, output generation, depicted in Alg.~\ref{algEuler}, requires the existence of Eulerian path in A-Bruijn graph $G(V,E)$.
This condition is tested by looking at all the vertices $v \in V$.
If all except the~initial and final vertices have an~equal number of input and output edges,
$\forall v \in V \setminus \{v_{0}, v_{n}\} : | v.out | = | v.in |$, the initial vertex satisfies $v_{0} \in V : | v.out | - | v.in | = 1$ condition,
and the final vertex $v_{n} \in V : | v.out | - | v.in | = -1$, the Eulerian path exists.

Due to the presence of repetitions, graph $G$ may contain many Eulerian paths, which means ambiguity of the~target sequence.
Therefore, the output generation constructs a~set of sub-strings of the target sequence called contigs.
The contig is an~Eulerian sub-path, as shown in Alg.~\ref{algEuler}.

\begin{algorithm}
  \caption{Algorithm to construct a~set of contigs from graph $G$.}
  \label{algEuler}
  \begin{algorithmic}
    \REQUIRE A-Bruijn graph $G$ with an~Eulerian path, $v_{0}$ is starting vertex
    \STATE  $contigs \gets \emptyset$ \COMMENT{output of assembler}
    \STATE $v \gets v_{0}$ \COMMENT{current vertex}
    \STATE $c \gets \emptyset$ \COMMENT{current contig}
    \LOOP
    \IF{$ | v.out | == 0$}
    \STATE $contigs.insert(c)$
    \RETURN $contigs$ \COMMENT{end of algorithm}
    \ENDIF
    \STATE $e \gets v.out[0]$ \COMMENT{current edge}
    \IF{$ | v.out | > 1$}
    \STATE $f \gets v.out[1]$ \COMMENT{second current edge}
    \IF{$ | v.out | = 2$ \AND ($isBridge(e)$ \OR $isBridge(f)$)}
    \IF{$isBridge(e)$}
    \STATE $e \gets f$
    \ENDIF
    \ELSE
    \STATE $contigs.insert(c)$  \COMMENT{$v$ is ambiguous}
    \STATE $c \gets \emptyset$
    \ENDIF
    \ENDIF
    \STATE $c.insert(e)$
    \STATE $v \gets e.target$
    \STATE $G.delete(e)$ \COMMENT{decreases edge's weight, if it achieves 0, remove $e$ from $G$}
    \ENDLOOP
  \end{algorithmic}
\end{algorithm}

The algorithm iteratively processes all vertices $V$ starting from $v_{0}$.
The current vertex $v$, if unambiguous, extends the current contig $c$,
otherwise, it starts the new contig.
A given vertex is unambiguous iff it has one or two output edges and, in the~case when exactly two output edges exist,
either is a~bridge (an~edge is a bridge if it has a~path leading from the~target vertex to the~source vertex).
This condition extends the test used in other existing assemblers, where ambiguity is set if the~vertex has more than one output edge.
Alg.~\ref{algEuler} reduces the~number of contigs in comparison with the~existing assemblers,
because the vertices with exactly two output edges, where one is a bridge,
do not create a~new contig.
Time complexity of the~presented algorithm is quadratic in function of the~number of edges, as is the~case for Fleury's algorithm \cite{cormen2001introduction}.

Alg.~\ref{algEuler} uses $isBridge(e)$ procedure that checks if edge $e$ is a~bridge, i.e. if it connects different strongly connected components.
This procedure tries to find a~path from target vertex of $e$ to source vertex of $e$ using depth first search \cite{cormen2001introduction}.

\subsection*{Assembly k-spectrum}

Given a string $S = s_{0}s_{1}...s_{G-1}$, $|S|=G$, let $S_{k}(i)$ be the sub-string $s_{i}...s_{i+k-1}$ of length $k \le G$.
The k-spectrum of $S$ is a set of all $S_{k}(i)$ for $0 \le i < G-k$.

\begin{figure}[!htb]
  \centering
 \scalebox{0.7}{\includegraphics{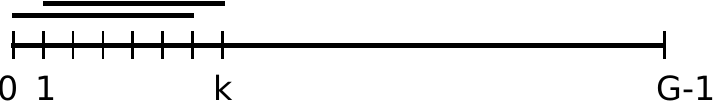}}
  \caption{K-spectrum construction.
    The reads are generated for coordinates $(0,k-1), (1,k), ..., (G-k,G-1)$, where $G$ is the input sequence length.}
  \label{figDeterministicFragments}
\end{figure}

The k-spectrum is idealized sequence assembler input, because all k-substrings without repetitions, errors and forward oriented
are available, as depicted in~Fig.~\ref{figDeterministicFragments}.

If k-spectrum is the input of presented algorithm, the edges weights inside whirls $w$ are expressed by Eq.~\ref{eqDetermWhirlsWeight},
where $n$ is SSR length, $d$ is motif length, $k$ is graph dimension.
Edges inside whirls have identical value $w$ for $n -k \equiv d -1 \pmod d$,
otherwise the weight is either $w$ or $w-1$, examples are shown in Tab.~\ref{tabWhirlsBruijn}.
In further we assume, for simplicity, that $n -k \equiv d -1 \pmod d$.

\begin{equation}
  w = \frac{\Delta}{d}, \mbox{~where } \Delta \equiv 0 \pmod d, \Delta = n-k+1
  \label{eqDetermWhirlsWeight}
\end{equation}

The presented algorithm is able to reconstruct SSR of any length from k-spectrum.
Eq.~\ref{eqDetermRepetiveLength} expresses repetitive sequence length from whirls parameters,
Tab.~\ref{tabDeterministicBruijnExamples} shows some examples.

\begin{equation}
  n = wd + k - 1, \mbox{~where } n - k \equiv  d - 1 \pmod d
  \label{eqDetermRepetiveLength}
\end{equation}

\begin{table}[!htb]
  \centering

  \begin{tabular}{|c|c|c|} \hline
    \scalebox{0.7}{\includegraphics{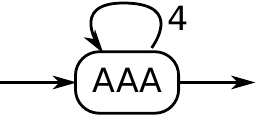}}
    & \scalebox{0.7}{\includegraphics{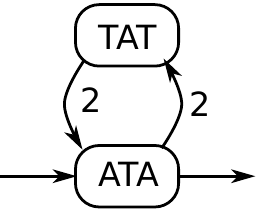}}
    & \scalebox{0.7}{\includegraphics{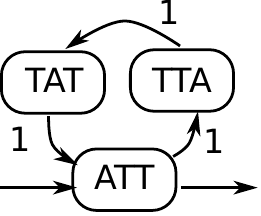}}
    \\
    \sSeq{AAAAAAA}
    & \sSeq{ATATATA}
    & \sSeq{ATTATT}
    \\ \hline
  \end{tabular}

  \caption{Examples of whirls in 4-dimensional A-Bruijn graph and SSR reconstructed using Eq.~\ref{eqDetermRepetiveLength}}
  \label{tabDeterministicBruijnExamples}
\end{table}

The ability of SSR reconstruction by the presented algorithm was checked on generated sequences.
The k-spectrum from these sequences was used as an~input for the presented computer program and the three existing genome assemblers based on de~Bruijn graph,
are described in Section~\ref{secMethods}.
Only the presented algorithm properly reconstructs the input, as depicted in Tab.~\ref{tabSpectrumReconstructionAssemblers}.

\subsection*{Assembly error-less uniformly distributed reads}

In this section, the randomly spread fragments are regarded to be the input.
All reads have the same length $L \ge k$,
the distribution of the initial positions is uniform,
the reads are properly oriented in the forward direction and the sequences have no errors.

This set of reads is closer to reality than k-spectrum, considered before,
all consecutive sub-strings are not required, the read length is not equal to graph dimension $k$.
The presented input is used to model the algorithm properties.

The reads are uniformly spread over the input genome sequence of length $G$, therefore
the probability $p$ of choosing fragment of length $k$ is $\frac{1}{G-L} \approx \frac{1}{G} \mbox{~for~} G \gg L$, where $L$ is the length of read.
The sequences assigned to the edges inside whirls are created from SSR, and the probability of choosing such a~sequence is
$p = \frac{\Delta}{dG}, \Delta = n-k+1$, where $d$ is the length of motif and $n$ the~length of SSR.
The weight of edge inside whirls $w$ is random variable with bi-nominal distribution,
because the reads are independent.
For $N$ reads, graph construction algorithm (Alg.~\ref{algGraphConstruct}) creates $N' = N(L-k+1)$ of sequences,
which allows to depict weight distribution $W$ as shown in Eq.~\ref{eqBinominal}.

\begin{equation}
  \begin{array}{l}
  W \sim B(N',p)\\
  \mbox{~where~} N' = N(L-k+1), p = \frac{\Delta}{dG}, \Delta = n-k+1
  \end{array}
  \label{eqBinominal}
\end{equation}

In real cases, the random variable $W$ depicted in Eq.~\ref{eqBinominal} could be estimated by Poisson distribution,
with $\lambda = N'p$, as depicted in Eq.~\ref{eqPoisson},
because $p \ll 1$ ($G \gg 1$) and the number of input fragments is big ($N \gg 1, L \ge k$).
The $c$ is sequence redundancy $c = \frac{LN}{G}$.
The parameter $c' = \frac{c(L-k+1)}{k}$ may be understood as edge redundancy,
the frequency of the sequence corresponding to the edge is in the output of Alg.~\ref{algGraphConstruct}.

\begin{equation}
  \begin{array}{l}
  W \sim \operatorname{Poisson} \left({\lambda}\right)\\
  \mbox{~where~} \lambda = \frac{c'\Delta}{d}, c' = \frac{c(L-k+1)}{k}, \Delta = n-k+1
  \end{array}
  \label{eqPoisson}
\end{equation}

\subsubsection*{Error estimation of edge weight normalization}

The edge's weight $w$ is normalized, then rounded, as depicted in Eq.~\ref{eqRounding},
to account for the reading coverage in the assembly algorithm.

The repetitive sequence length $\hat{n}$ could be estimated from A-Bruijn graph whirls parameters by using Eq.~\ref{eqRepLengthEstimation},
where $w'$ is normalized edge's weight, $k$ is graph dimension, $d$ is motif length.
This relation is similar to Eq.~\ref{eqDetermWhirlsWeight} defined for k-spectrum input.

\begin{equation}
  \hat{n} = w'd + k - 1, \mbox{~where~} n - k \equiv d - 1 \pmod d
  \label{eqRepLengthEstimation}
\end{equation}

\begin{figure}[!htb]

 \scalebox{0.4}{\includegraphics{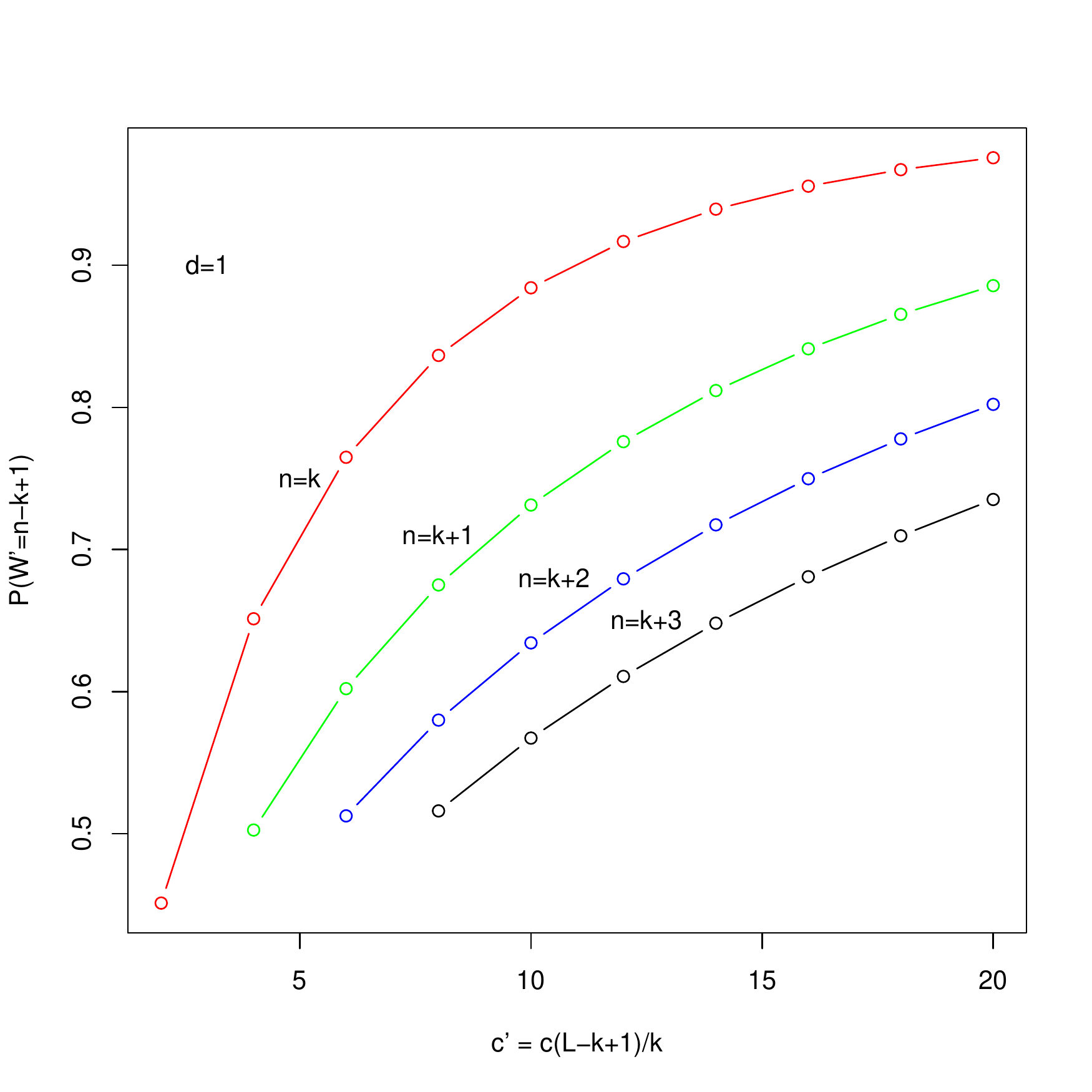}}
  \caption{The probability of proper SSR length estimation in the function of coverage.
    The Eq.~\ref{eqCorrectProbabilityValue} was used for single symbol repetitions.}
  \label{figUniformProbNumbers}
\end{figure}

The probability of getting $w'$ that allows to determine accurately the length of SSR is depicted in~Eq.~\ref{eqCorrectProbabilityValue},
where $\Phi_{P}(x, \lambda)$ is cumulative distribution function for Poisson distribution with parameter $\lambda$.
This relation is depicted in~Fig.~\ref{figUniformProbNumbers}.

\begin{equation}
  \begin{array}{l}
  P(w'=\frac{n-k+1}{d}) = \Phi_{P}(\lambda + \frac{c'}{2}, \lambda) - \Phi_{P}(\lambda - \frac{c'}{2}, \lambda)\\
  \mbox{~where~} \lambda, c' \mbox{~are defined in Eq.~\ref{eqPoisson}}
  \end{array}
  \label{eqCorrectProbabilityValue}
\end{equation}

The probability mass function of a~variable with Poisson distribution is asymmetrical:
it is high on the left and skewed towards the right,
therefore the rounding error made at the right end of the interval is larger than at the left end.
It advises not to use rounding to the nearest integer (as depicted in Eq.~\ref{eqRounding}).

\begin{figure}[!htb]

 \scalebox{0.4}{\includegraphics{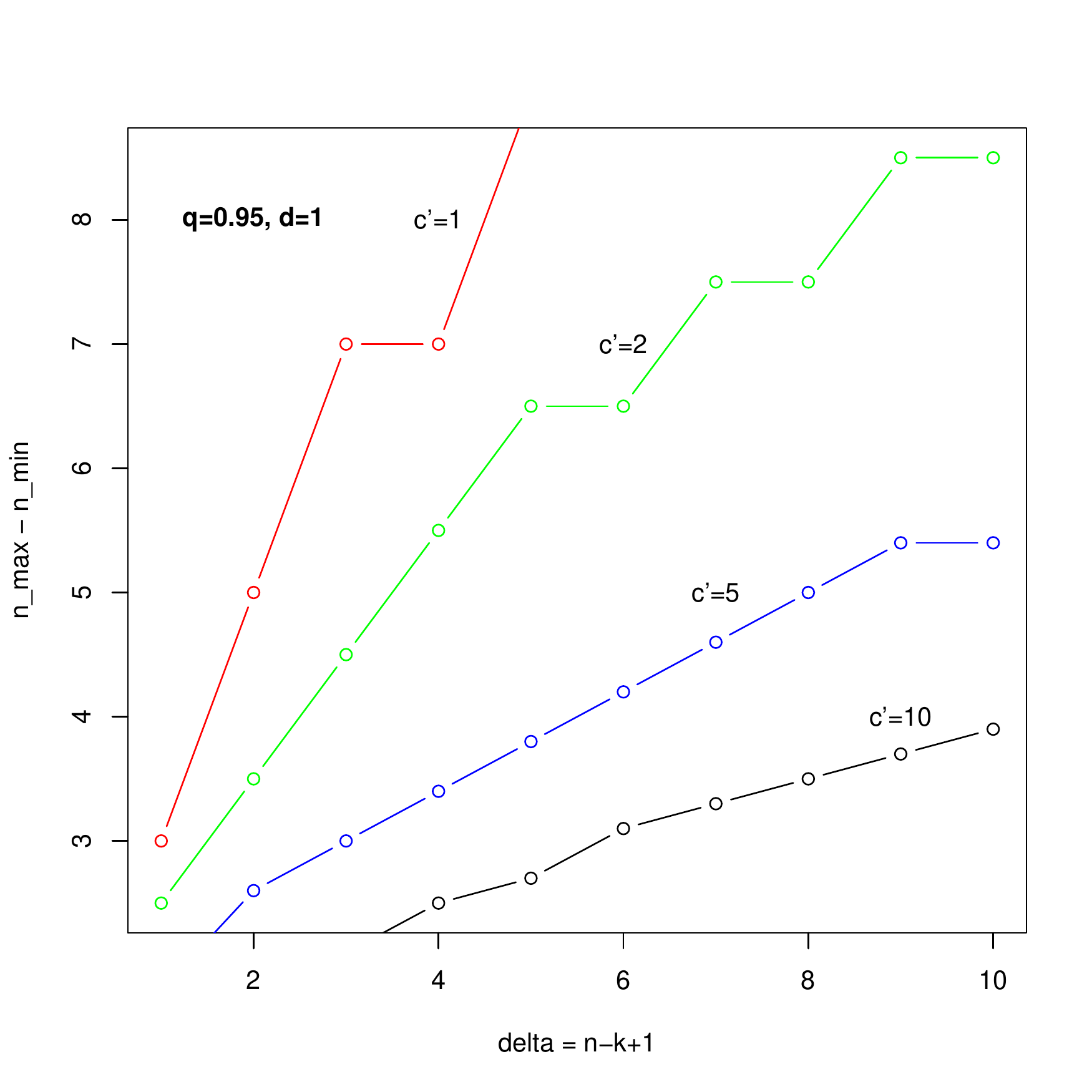}}
 \scalebox{0.4}{\includegraphics{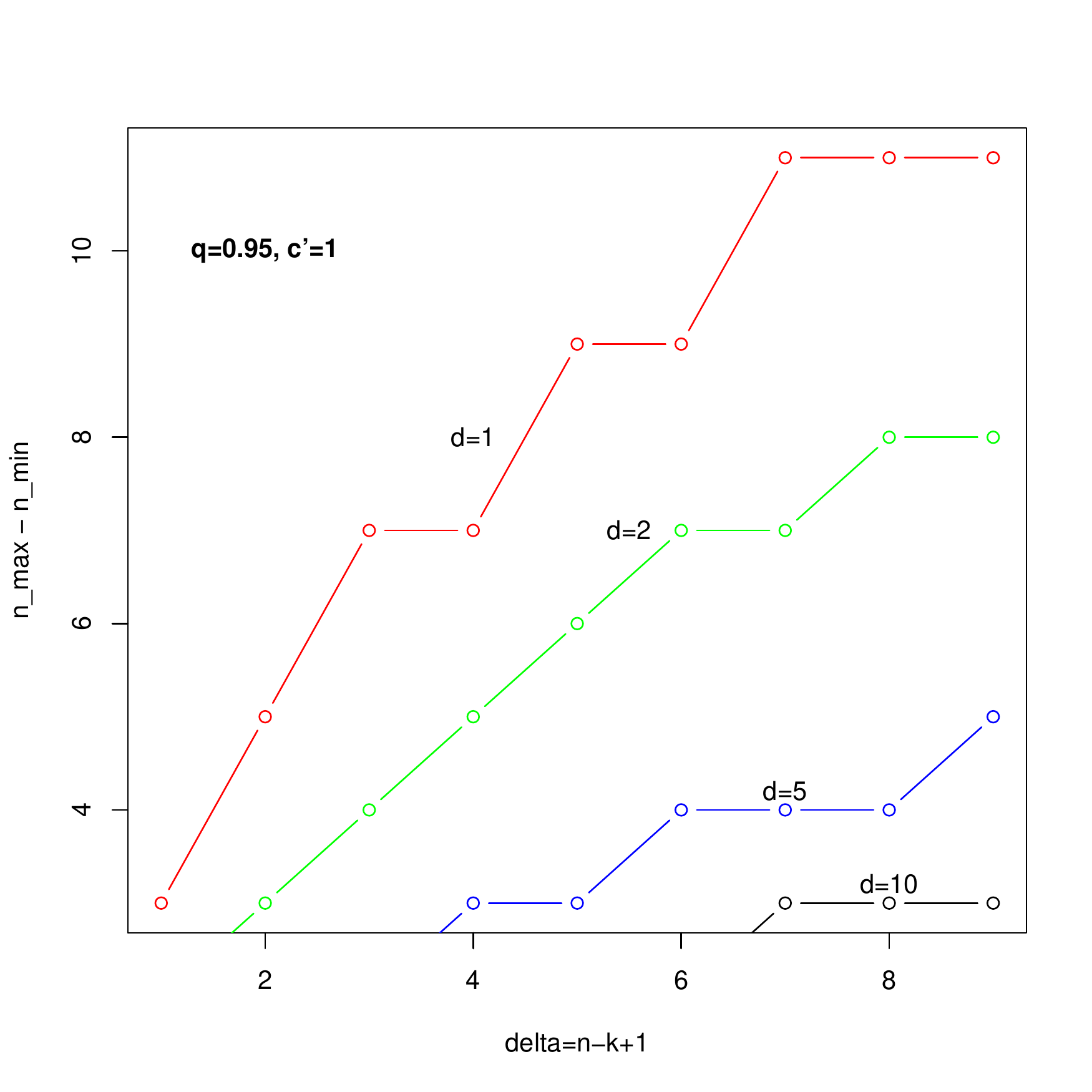}}

  \caption{The interval size ($\hat{n}_{max} - \hat{n}_{min}$ from Eq.~\ref{eqRepLengthInterval}) of estimated SSR length for $0.95$ confidence level.}
  \label{figUniformDeltaByLengthP95}
\end{figure}

The repetitive sequence length, estimated by Eq.~\ref{eqRepLengthEstimation},
is inside the interval defined by Eq.~\ref{eqRepLengthInterval},
where $0 \le q \le 1$ is a~defined level of confidence,
$\Phi^{-1}_{P}(q, \lambda)$ is an~inverse cumulative distribution function for Poisson distribution with parameter $\lambda$,
$\hat{n}_{max}$ is an~upper boundary of repetitive sequence length,
$\hat{n}_{min}$ is a~lower boundary of repetitive sequence length.
The intervals are depicted in Fig.~\ref{figUniformDeltaByLengthP95}.

\begin{equation}
  \left.
  \begin{array}{ll}
    \hat{n}_{max} &= \frac{1}{c'} \Phi^{-1}_{P}(\frac{1+q}{2}, \lambda)\\
    \hat{n}_{min} &= \frac{1}{c'} \Phi^{-1}_{P}(\frac{1-q}{2}, \lambda)
  \end{array}
  \right\}
  \mbox{~where~} \lambda, c' \mbox{~are defined in Eq.~\ref{eqPoisson}}
  \label{eqRepLengthInterval}
\end{equation}

\subsubsection*{Calculating the required sequencing coverage}

The SSR length estimation requires high coverage, as depicted in Fig.~\ref{figUniformProbNumbers},
where $P(w'= \frac{n-k+1}{d}) \ge 0.9$ for $c' > 10$.
For high coverage, the Poisson distribution of edge's weight, Eq.~\ref{eqPoisson}, can be approximated by Normal distribution,
as depicted in Eq.~\ref{eqNormal}, where $\mu$ is a~mean, $\sigma$ is a~standard deviation.

\begin{equation}
  W' \sim \mathcal{N}(\mu, \sigma) \mbox{~where~} \mu = \frac{\Delta}{d}, \sigma = \sqrt{\frac{\Delta}{d}}, \Delta = n-k+1
  \label{eqNormal}
\end{equation}

\begin{figure}[!htb]
 \scalebox{0.4}{\includegraphics{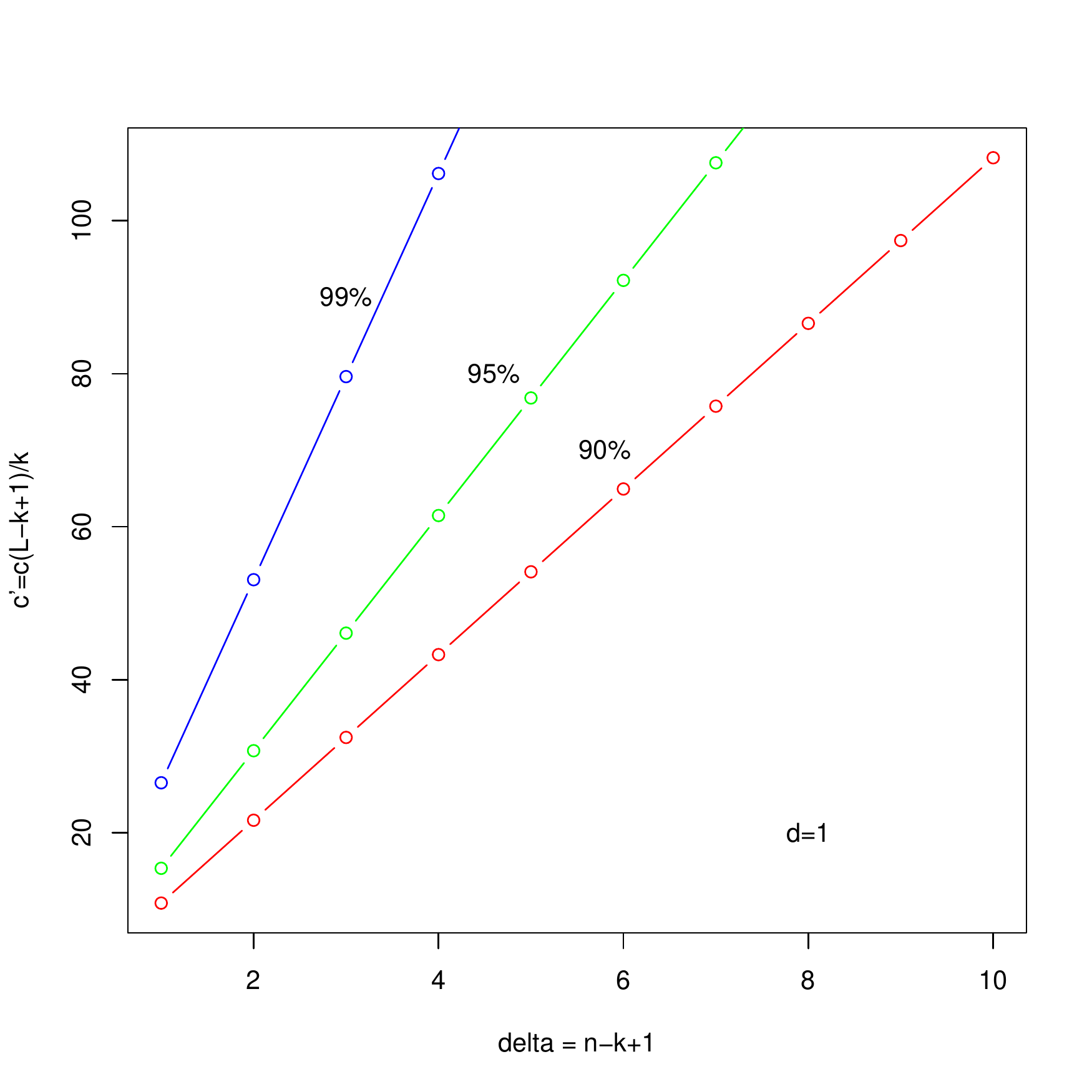}}
 \scalebox{0.4}{\includegraphics{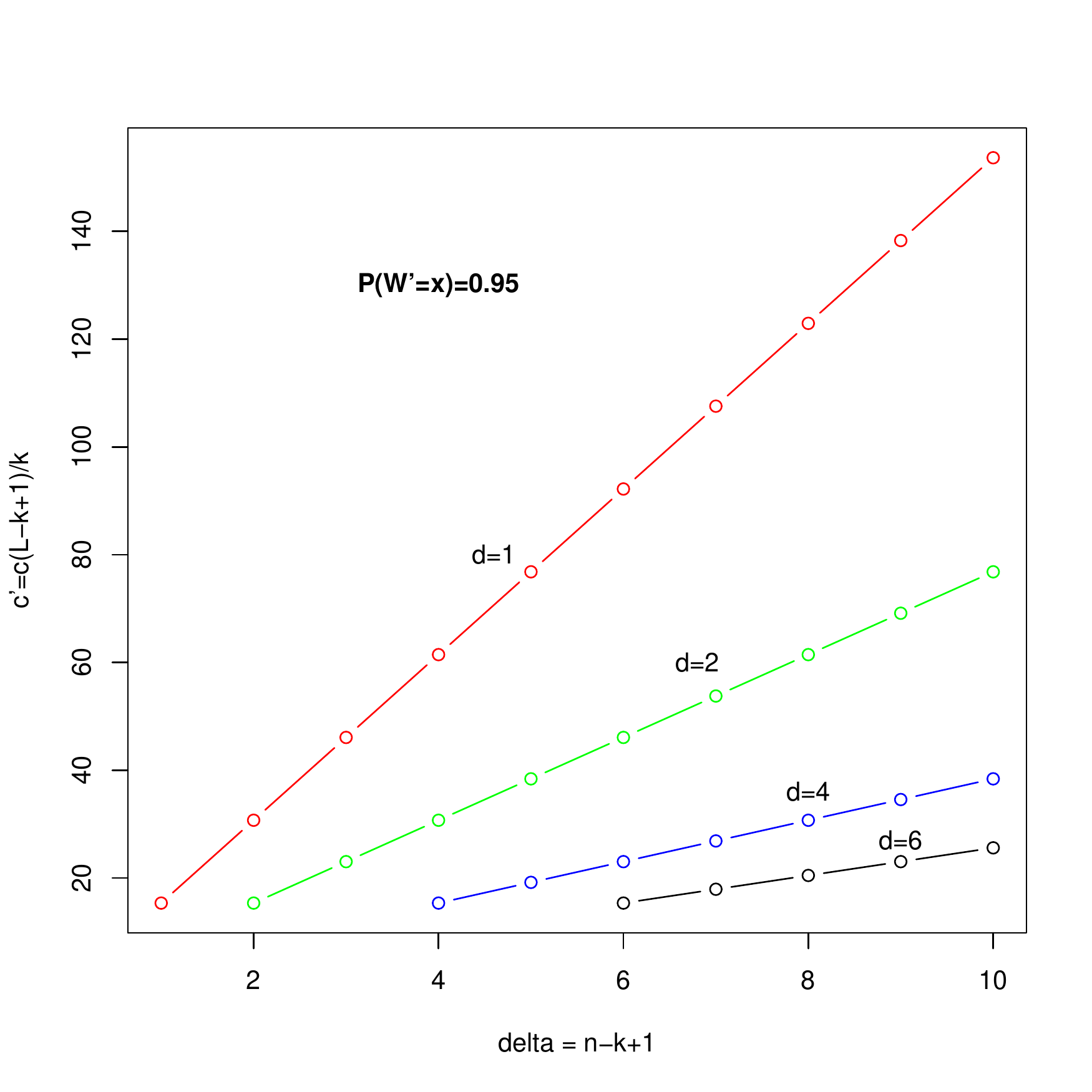}}
  \label{figUniforRedundancyByLength}
  \caption{The required edge coverage $c'$ for high values of confidence level in function of SSR length.}
\end{figure}

In this case the required coverage is linearly dependent on the repetitive sequence length $n$, as depicted in Eq.~\ref{eqCoverageNormal}
and Fig.~\ref{figUniforRedundancyByLength},
where $\Phi_{N}^{-1}(q)$ is the inverse cumulative distribution function for standard normal distribution ($\mu = 0$, $\sigma = 1$),
$q$ is required confidence level.

\begin{equation}
  c' = (2 \Phi_{N}^{-1}(\frac{1+q}{2}))^2 \frac{\Delta k}{d} \mbox{~where~} \Delta = n-k+1
  \label{eqCoverageNormal}
\end{equation}

This model is useful in planning genome assembly experiments,
for example, to achieve estimation of the length of the repetitive sequences of $d=5$ and $n=150$ with 95\% confidence interval,
when the average read length is $L=80$, the de Bruijn graph dimension $k=50$,
we should use $c' \ge 307$, therefore $c \ge 495$.
If the $30\times$ coverage is used ($c = 30, c'=\frac{c(L-k+1)}{k}=31$) the length of repetitive sequence of $d=5$ and $n=150$
is in interval $\hat{n} \in (146,154)$ with 95\% probability.

\section{Methods}
\label{secMethods}

\subsection*{Reconstruct generated SSR}

The numerical experiments prove the ability to reconstruct SSR properly by the presented algorithm.
The $18$ different input sequences were generated by Python script\footnote{available at project homepage}.
These sequences can be described by patterns $A$, $AR$, $ARB$, where the $A$, $B$ and $R$ are sequences of length $1000$ and $10000$,
$A$ and $B$ are unique without SSR, $R$ is SSR with a~motif of length $d = 1,2,4,10$.
The k-spectrum for each sequence mentioned above were generated and used as input for presented algorithm,
ABySS \cite{simpson2009abyss}, ver. 1.3.7 genome assembler, MIRA\cite{chevreux1999genome}, ver. 4.0 and Velvet\cite{zerbino2008velvet} ver. 1.2.10.
Each input was performed 3 times for each assembler, to eliminate the randomness.
Tab.~\ref{tabSpectrumReconstructionAssemblers} summarizes the results:
the presented approach properly assembles SSR longer than graph dimension,
the other assemblers are unable to assemble the repetitive sequences.

\begin{table}[!htb]
  \begin{tabular}{|l|c|c|c|c|} \hline
    sequence             & presented & ABySS & MIRA & Velvet \\ \hline
    $A$                  & + & + & + & + \\
    $AR$                 & + & - & - & - \\
    $ARB$                & + & - & - & - \\
    \hline
  \end{tabular}
  \caption{The successful reconstruction of generated sequence from its k-spectrum.
    $A, B$ are long, unique, non-repetitive sequences,
    $R$ is long SSR, $|A| = |B| = |R| = 1000$ or $|A| = |B| = |R| = 10000$, motif length is $1,2,4$ or $10$.}
  \label{tabSpectrumReconstructionAssemblers}
\end{table}

\subsection*{Assembly k-spectrum generated from model genomes}

The existing genome sequences of model species: \textit{Escherichia coli} (536 genome, GenBank NC\_008253),
\textit{Saccharomyces cerevisiae} (S288c genome, version R64, GenBank NC\_001133 ...NC\_001148, NC\_001224),
\textit{Arabidopsis thialina} (TAIR9, The Arabidopsis Information Resource \url{http://arabidopsis.org}),
and \textit{Homo sapiens} (GRCh37, release 75, \url{http://www.ensembl.org})
were investigated to check the regions properly assembled by the presented algorithm.

Firstly, the k-spectrum, where $k=51$ was generated for each chromosome;
secondly, the assembler was used to reconstruct the sequence.
The  number of places where the presented algorithm properly assembles the sequence and a~typical algorithm fails was counted.
The results are depicted in~Tab.~\ref{tabNoAlgSuccess}.

The presented algorithm achieves approximately 5\% less contig numbers than other assemblers.
Tab.~\ref{tabSSRFromRealData} and Tab.~\ref{tabSSRFromHumanRealData} depicts examples of properly reconstructed SSR.

\begin{table}[!htb]
  \begin{tabular}{|l|l|} \hline
    name & no. of places \\ \hline
    e.coli, \textit{Escherichia coli} & 53 \\
    yeast, \textit{Saccharomyces cerevisiae} & 301 \\
    arabidopsis, \textit{Arabidopsis thialina} & 2911 \\
    human, \textit{Homo sapiens}, chr.\ XIII -- XXII & 15565 \\ \hline
  \end{tabular}
  \caption{Number of places where our algorithm works properly and other sequence assemblers fail. The reads were generated from model genomes.}
  \label{tabNoAlgSuccess}
\end{table}

\begin{table}[!htb]
  \begin{tabular}{|l|c|c|} \hline
    position (index) & motif & length \\ \hline
    e.coli, 2066694 & ACAGATAC & 80 \\ \hline
    yeast, chr.II, 464020 & GTT & 83 \\ \hline
    yeast, chr.IV, 778759 & TTA & 71 \\ \hline
    yeast, chr.VII, 280054 & GAGGTTGCTGTT & 110 \\ \hline
    yeast, chr.XIII, 86952 & TTA & 118 \\ \hline
    arabid., chr.I, 17882880 & ATC & 125 \\ \hline
    arabid., chr.I, 18486514 & TGTA & 135 \\ \hline
    arabid., chr.II, 2758565 & TTCTATG & 130 \\ \hline
    arabid., chr.II, 5401576 & TTA & 145 \\ \hline
    arabid., chr.II, 16723214 & CAGTCT & 135 \\ \hline
    arabid., chr.III, 1818686 & TAA & 71 \\ \hline
    arabid., chr.III, 6375808 & ATGGGG & 86 \\ \hline
    arabid., chr.IV, 2488534 & AAGACGAAGAAG & 90 \\ \hline
    arabid., chr.V, 7614543 & TACA & 82 \\ \hline
    arabid., chr.V, 20395151 & TAA & 180 \\ \hline
    arabid., chr.V, 24022835 & TCC & 140 \\ \hline
  \end{tabular}

  \caption{SSR properly assembled from k-spectrum generated from e.coli (\textit{Escherichia coli}),
    yeast (\textit{Saccharomyces cerevisiae}), arabidopsis (\textit{Arabidopsis thialina}) genomes, where $k = 51$.
    The table depicts SSR longer than $70$ of motif length $d \le 12$.}
  \label{tabSSRFromRealData}

\end{table}

\begin{table}[!htb]
  \begin{tabular}{|l|c|c|c|} \hline
    chr. & index & motif & length \\ \hline
    XIII  & 29027164  & TTTCC & 135 \\ \hline
    XIII  & 49892614  & GGAAAG & 167 \\ \hline
    XIII  & 44716268  & CTCGG & 180 \\ \hline
    XIII  & 102813925 & AGA & 152 \\ \hline
    XV    & 65438307 & TAAATATATATA & 161 \\ \hline
    XV    & 69970764 & CTTTC & 105 \\ \hline
    XVII  &  5185392 & CGCGCTCCCTC & 109 \\ \hline
    XVII  & 20460059 & TCCCTC & 149 \\ \hline
    XVII  &  4365394 & TCCA & 533 \\ \hline
    XVII  & 77867600 & ATC & 239 \\ \hline
    XVII  & 78639134 & TTCCT & 170 \\ \hline
    XVIII & 47105376 & AGAGGG & 233 \\ \hline
    XVIII & 49836110 & ATATATATTTCT & 112 \\ \hline
    XVIII & 62056751 & CTTTC & 125 \\ \hline
    XIX   & 53422228 & CTCCCT & 194 \\ \hline
    XIX   & 57993815 & CTCTCCC & 120 \\ \hline
    XXI   & 40955746 & CCTT & 254 \\ \hline
    XXII  & 16288601 & CGGCGTGCGCGTG & 102 \\ \hline
    XXII  & 27691662 & ATGG & 102 \\ \hline
  \end{tabular}

  \caption{SSR properly assembled from k-spectrum generated from human genome (\textit{Homo sapiens}), chromosomes XIII -- XXII, where $k = 51$.
    The table depicts SSR longer than $100$ of motif length $d \le 12$.}

  \label{tabSSRFromHumanRealData}

\end{table}

\section{Discussion and conclusion}
\label{secDiscussion}

The presented algorithm uses the short read data more efficiently than other computer programs when high coverage is available.
It is able to assemble properly some repetitive sequences, and achieve 5\% better contig size, if used on k-spectrum generated from model genomes.

This assembler is able to use reads of different lengths, if the length is greater than graph dimension $k$.
If you are using collections of readings of different lengths, e.g. from different experiments,
the fragment length $L$ should be replaced by arithmetic mean $\bar{L}$
in Eq.~\ref{eqBinominal}, Eq.~\ref{eqPoisson}, Eq.~\ref{eqCorrectProbabilityValue} and Eq.~\ref{eqRepLengthInterval}.

The existing genome drafts, especially for plant genomes, are not fully assembled, inter alia, due to a~large number of SSR.
The presented algorithm could reduce gaps in the existing data.
The procedure includes finding the repetitive sequences in contigs ends.
If two and only two contigs have the same SSR at the end, they could be connected to create a~single contig.
The SSR length is estimated by Eq.~\ref{eqRepLengthEstimation} with accuracy expressed in Eq.~\ref{eqRepLengthInterval}.

To obtain the length of repetitive sequence with a~big confidence level, higher than typical coverage should be applied ($c > 100$).
If the required coverage is beyond the project funds, it should be considered in future genome sequencing projects,
as the cost decreases exponentially.

The mathematical model proposed in this paper is the upper limit of the calculation accuracy,
because the distribution of read start positions deviates from uniformity and contains sequencing errors.
The big coverage required by the presented approach, when normal distribution describes the edge's weight,
permit the underlying distribution different from uniform.
However, the special sequences, underestimated and overestimated with the sequencing technology should be considered.
We plan to review the technology to find such sequences and take this into account in the next version of software.

The sequencing errors are corrected in the presented approach by edge normalization process (Eq.~\ref{eqRounding}).
The other error awareness techniques are considered in future versions of presented software:
incorporate sequence quality into assembly algorithm
and correction of systematic errors created by next-generation sequencers.
Moreover, the improper assembly output could be corrected when information of k-mer position generated from read is used \cite{ronen2012sequel}.

The key improvement in assembly results, especially for de Bruijn graph based solutions,
is the usage of the new experimental opportunity, called mate pairs.
The sequencers are able to read the pairs of sequences between which the genomic distance is well estimated.
The mate pairs could link contigs into scaffolds, and are used
either as post-processing steps \cite{butler2008allpaths,zerbino2008velvet,piotrowski2013new}
or as internal assembler process where connections are incorporated into the graph structure \cite{medvedev2011paired}.
Mate pairs, in theory, allow to properly assemble the sequences of length equal to
the sum of lengths of known sequences on both ends and length of the distance.
It significantly increases the effective length of reads, because the distance could be long.
The mate pairs were successfully used to de novo assemble highly repetitive telomeric regions \cite{bresler2012telescoper}.

The presented algorithm currently does not use mate pair data;
however, the integration of such data is one of the most important tasks in the next version of software.

The presented algorithm assumes forward orientation of all readings,
i.e. the reads come from only one DNA strand.
In real data both coding and complementary DNA strand provide reads.
The further version of computer program will include the sequence and complementary in node representation,
similarly to the Velvet assembler \cite{zerbino2008velvet}.

In conclusion, the proposed solution better assembles short sequence repeats than other sequence assemblers.
The proposed mathematical model estimates the coverage to achieve the required level of confidence.
The length of properly reconstructed SSR linearly depends on genome coverage.

\section*{Acknowledgement}
This work was supported by the statutory research of Institute of Electronic Systems of Warsaw University of Technology.

\bibliographystyle{abbrv}
\bibliography{rnowak_ssr_assembly}

\begin{thebibliography}{10}
\expandafter\ifx\csname url\endcsname\relax
  \def\url#1{\texttt{#1}}\fi
\expandafter\ifx\csname urlprefix\endcsname\relax\def\urlprefix{URL }\fi
\expandafter\ifx\csname href\endcsname\relax
  \def\href#1#2{#2} \def\path#1{#1}\fi

\bibitem{shendure2008next}
J.~Shendure, H.~Ji, Next-generation dna sequencing, Nature biotechnology
  26~(10) (2008) 1135--1145.

\bibitem{pagani2012genomes}
I.~Pagani, K.~Liolios, J.~Jansson, I.-M.~A. Chen, T.~Smirnova, B.~Nosrat, V.~M.
  Markowitz, N.~C. Kyrpides, The genomes online database (gold) v. 4: status of
  genomic and metagenomic projects and their associated metadata, Nucleic acids
  research 40~(D1) (2012) D571--D579.

\bibitem{pevzner2001eulerian}
P.~A. Pevzner, H.~Tang, M.~S. Waterman, An eulerian path approach to dna
  fragment assembly, Proceedings of the National Academy of Sciences 98~(17)
  (2001) 9748--9753.

\bibitem{myers2005fragment}
E.~W. Myers, The fragment assembly string graph, Bioinformatics 21~(suppl 2)
  (2005) ii79--ii85.

\bibitem{miller2010assembly}
J.~R. Miller, S.~Koren, G.~Sutton, Assembly algorithms for next-generation
  sequencing data, Genomics 95~(6) (2010) 315--327.

\bibitem{zhang2011practical}
W.~Zhang, J.~Chen, Y.~Yang, Y.~Tang, J.~Shang, B.~Shen, A practical comparison
  of de novo genome assembly software tools for next-generation sequencing
  technologies, PloS one 6~(3) (2011) e17915.

\bibitem{earl2011assemblathon}
D.~Earl, K.~Bradnam, J.~S. John, A.~Darling, D.~Lin, J.~Fass, H.~O.~K. Yu,
  V.~Buffalo, D.~R. Zerbino, M.~Diekhans, et~al., Assemblathon 1: A competitive
  assessment of de novo short read assembly methods, Genome research 21~(12)
  (2011) 2224--2241.

\bibitem{bradnam2013assemblathon}
K.~R. Bradnam, J.~N. Fass, A.~Alexandrov, P.~Baranay, M.~Bechner, I.~Birol,
  S.~Boisvert, J.~A. Chapman, G.~Chapuis, R.~Chikhi, et~al., Assemblathon 2:
  evaluating de novo methods of genome assembly in three vertebrate species,
  GigaScience 2~(1) (2013) 1--31.

\bibitem{salzberg2012gage}
S.~L. Salzberg, A.~M. Phillippy, A.~Zimin, D.~Puiu, T.~Magoc, S.~Koren, T.~J.
  Treangen, M.~C. Schatz, A.~L. Delcher, M.~Roberts, et~al., Gage: A critical
  evaluation of genome assemblies and assembly algorithms, Genome research
  22~(3) (2012) 557--567.

\bibitem{kingsford2010assembly}
C.~Kingsford, M.~C. Schatz, M.~Pop, Assembly complexity of prokaryotic genomes
  using short reads, BMC bioinformatics 11~(1) (2010) 21.

\bibitem{cox1997characteristic}
R.~Cox, S.~M. Mirkin, Characteristic enrichment of dna repeats in different
  genomes, Proceedings of the National Academy of Sciences 94~(10) (1997)
  5237--5242.

\bibitem{van1998short}
A.~van Belkum, S.~Scherer, L.~van Alphen, H.~Verbrugh, Short-sequence dna
  repeats in prokaryotic genomes, Microbiology and Molecular Biology Reviews
  62~(2) (1998) 275--293.

\bibitem{cao2013inferring}
M.~D. Cao, E.~Tasker, K.~Willadsen, M.~Imelfort, S.~Vishwanathan,
  S.~Sureshkumar, S.~Balasubramanian, M.~Bod{\'e}n, Inferring short tandem
  repeat variation from paired-end short reads, Nucleic acids research (2013)
  gkt1313.

\bibitem{xie2009cnv}
C.~Xie, M.~T. Tammi, Cnv-seq, a new method to detect copy number variation
  using high-throughput sequencing, BMC bioinformatics 10~(1) (2009) 80.

\bibitem{yoon2009sensitive}
S.~Yoon, Z.~Xuan, V.~Makarov, K.~Ye, J.~Sebat, Sensitive and accurate detection
  of copy number variants using read depth of coverage, Genome research 19~(9)
  (2009) 1586--1592.

\bibitem{chaisson2008short}
M.~J. Chaisson, P.~A. Pevzner, Short read fragment assembly of bacterial
  genomes, Genome research 18~(2) (2008) 324--330.

\bibitem{pevzner2004novo}
P.~Pevzner, H.~Tang, G.~Tesler, De novo repeat classification and fragment
  assembly, Genome Research 14~(9) (2004) 1786--1796.

\bibitem{cormen2001introduction}
T.~Cormen, C.~Leiserson, R.~Rivest, C.~Stein, Introduction to algorithms, The
  MIT press, 2001.

\bibitem{simpson2009abyss}
J.~T. Simpson, K.~Wong, S.~D. Jackman, J.~E. Schein, S.~J. Jones, {\.I}.~Birol,
  Abyss: a parallel assembler for short read sequence data, Genome research
  19~(6) (2009) 1117--1123.

\bibitem{chevreux1999genome}
B.~Chevreux, T.~Wetter, S.~Suhai, Genome sequence assembly using trace signals
  and additional sequence information., in: German Conference on
  Bioinformatics, 1999, pp. 45--56.

\bibitem{zerbino2008velvet}
D.~R. Zerbino, E.~Birney, Velvet: algorithms for de novo short read assembly
  using de bruijn graphs, Genome research 18~(5) (2008) 821--829.

\bibitem{ronen2012sequel}
R.~Ronen, C.~Boucher, H.~Chitsaz, P.~Pevzner, Sequel: improving the accuracy of
  genome assemblies, Bioinformatics 28~(12) (2012) i188--i196.

\bibitem{butler2008allpaths}
J.~Butler, I.~MacCallum, M.~Kleber, I.~A. Shlyakhter, M.~K. Belmonte, E.~S.
  Lander, C.~Nusbaum, D.~B. Jaffe, Allpaths: de novo assembly of whole-genome
  shotgun microreads, Genome research 18~(5) (2008) 810--820.

\bibitem{piotrowski2013new}
P.~Piotrowski, R.~Nowak, New tool to combine contigs by usage of paired-end
  tags, in: Photonics Applications in Astronomy, Communications, Industry, and
  High-Energy Physics Experiments 2013, International Society for Optics and
  Photonics, 2013, pp. 890318--890318.

\bibitem{medvedev2011paired}
P.~Medvedev, S.~Pham, M.~Chaisson, G.~Tesler, P.~Pevzner, Paired de bruijn
  graphs: a novel approach for incorporating mate pair information into genome
  assemblers, Journal of Computational Biology 18~(11) (2011) 1625--1634.

\bibitem{bresler2012telescoper}
M.~Bresler, S.~Sheehan, A.~H. Chan, Y.~S. Song, Telescoper: de novo assembly of
  highly repetitive regions, Bioinformatics 28~(18) (2012) i311--i317.

\end{thebibliography}

\end{document}